\documentclass[prl,twocolumn,showpacs]{revtex4}
\usepackage{amsmath,amsthm,amsfonts,amscd,mathrsfs,pifont,bm}
\usepackage{eucal}

\renewcommand{\atop}[2]{\genfrac{}{}{0pt}{}{#1}{#2}}

\begin{document}

\title  {Integrable theory of quantum transport in chaotic cavities}

\author {Vladimir Al.~Osipov$\,{}^{1,2}$ and Eugene Kanzieper$\,{}^1$}

\affiliation
       {
       $\,{}^1$Department of Applied Mathematics, H.I.T.---Holon Institute of
       Technology,
       Holon 58102, Israel\\
       ${}^2$Fachbereich Physik, Universit\"at Duisburg-Essen, D-47057 Duisburg, Germany
}
\date   {June 17, 2008}

\begin  {abstract}
The problem of quantum transport in chaotic cavities with broken
time-reversal symmetry is shown to be completely integrable in the
universal limit. This observation is utilised to determine the
cumulants and the distribution function of conductance for a cavity
with ideal leads supporting an arbitrary number $n$ of propagating
modes. Expressed in terms of solutions to the fifth Painlev\'e
transcendent and/or the Toda lattice equation, the conductance
distribution is further analysed in the large-$n$ limit that reveals
long exponential tails in the otherwise Gaussian curve.
\end{abstract}

 \maketitle

{\it Introduction.}---The low temperature electronic conduction
through a cavity exhibiting chaotic classical dynamics is governed
by quantum phase-coherence effects \cite{AMPZJ-1995,I-2002}. In the
absence of electron-electron interactions \cite{R-01,B-1997,A-2000},
the most comprehensive theoretical framework by which the phase
coherent electron transport can be explored is provided by the
scattering ${\mathcal S}$-matrix approach pioneered by Landauer
\cite{LFLB-1957}. There exist two different, though mutually
overlapping, scattering-matrix descriptions \cite{LW-1991} of
quantum transport.

A semiclassical formulation \cite{R-2000} of the ${\mathcal
S}$-matrix approach is tailor-made to the analysis of
energy-averaged charge conduction \cite{R-02} through an individual
cavity. Representing quantum transport observables (such as
conductance, shot-noise power, transferred charge etc.) in terms of
classical trajectories connecting the leads attached to a cavity,
the semiclassical approach \cite{SSF} efficiently accounts for
system-specific features \cite{AL-1996} of the quantum transport.
Besides, it also covers the long-time scale universal transport
regime \cite{RS-2002} emerging in the limit \cite{R-00} $\tau_{\rm
D} \gg \tau_{\rm E}$, where  $\tau_{\rm D}$ is the average electron
dwell time and $\tau_{\rm E}$ is the Ehrenfest time (the time scale
where quantum effects set in).

The latter {\it universal regime} \cite{R-03} can alternatively be
studied within a stochastic approach \cite{B-1997,B-1995} based on a
random matrix description \cite{BGS-1984} of electron dynamics in a
cavity. Modelling a single electron Hamiltonian by an $M \times M$
random matrix ${\mathcal H}$ of proper symmetry, the stochastic
approach starts with the Hamiltonian $H_{\rm tot}$ of the total
system comprised by the cavity and the leads:
\begin{eqnarray}
\label{ham}
    H_{\rm tot} &=& \sum_{k,\ell=1}^M {\bm \psi}_k^\dagger {\mathcal H}_{k\ell}
    {\bm \psi}_\ell +  \sum_{\alpha=1}^{N_{\rm L}+N_{\rm R}}
    {\bm \chi}_{\alpha}^\dagger \varepsilon_F {\bm \chi}_{\alpha}\nonumber \\
    &+& \sum_{k=1}^M \sum_{\alpha=1}^{N_{\rm L}+N_{\rm R}} \left(
    {\bm \psi}_k^\dagger
    {\mathcal W}_{k \alpha} {\bm \chi}_{\alpha} +
    {\bm \chi}_{\alpha}^\dagger {\mathcal W}_{k \alpha}^* {\bm \psi}_k
    \right).
\end{eqnarray}
Here, ${\bm \psi}_k$ and ${\bm \chi}_\alpha$ are the annihilation
operators of electrons in the cavity and in the leads, respectively.
Indices $k$ and $\ell$ enumerate electron states in the cavity:
$1\le k,\ell \le M$, with $M\rightarrow \infty$. Index $\alpha$
counts propagating modes in the left ($1\le \alpha \le N_{\rm L}$)
and the right ($N_{\rm L}+1 \le \alpha \le N$) lead. The
$M\times N$ matrix ${\mathcal W}$ describes the coupling of electron
states with the Fermi energy ${\varepsilon_F}$ in the cavity to
those in the leads; $N=N_{\rm L}+N_{\rm R}$ is the total number of
propagating modes (channels). Since in Landauer-type theories the
transport observables are expressed in terms of the $N\times N$
scattering matrix \cite{A-2000}
\begin{eqnarray}
\label{sm}
    {\mathcal S}(\varepsilon_F) = \openone - 2i \pi  {\mathcal W}^\dagger
    ({\varepsilon}_F - {\mathcal H} + i \pi {\mathcal W}
    {\mathcal W}^\dagger)^{-1} {\mathcal W},
\end{eqnarray}
the knowledge of its distribution is central to the stochastic
approach. (Two such observables -- the conductance\linebreak $G =
{\rm tr\,} ({\mathcal C}_1 {\mathcal S} {\mathcal C}_2 {\mathcal
S}^\dagger)$ and the shot noise power $P = {\rm tr\,} ({\mathcal
C}_1 {\mathcal S} {\mathcal C}_2 {\mathcal S}^\dagger) - {\rm tr\,}
({\mathcal C}_1 {\mathcal S} {\mathcal C}_2 {\mathcal S}^\dagger)^2$
measured in proper dimensionless units \cite{B-1997} -- are of most
interest. Here, ${\mathcal C}_1={\rm diag}(\openone_{N_{\rm L}},
0_{N_{\rm R}})$ and ${\mathcal C}_2={\rm diag}(0_{N_{\rm L}},
\openone_{N_{\rm R}})$ are the projection matrices.)

For random matrices ${\mathcal H}$ drawn from rotationally
invariant Gaussian ensembles \cite{M-2004}, the distribution of
${\mathcal S}(\varepsilon_F)$ is described \cite{B-1995} by the
Poisson kernel \cite{R-05,H-1963,MB-1999}
\begin{eqnarray}
\label{pk}
    P({\mathcal S}) \propto \left[
    {\rm det}\left(
        \openone -  \bar{\mathcal S} {\mathcal S}^\dagger\right)
        {\rm det}\left(
        \openone -   {\mathcal S} \bar{\mathcal S}^\dagger\right)
    \right]^{ \beta/2 -1  -  \beta N/2}.
\end{eqnarray}
Here, $\beta$ is the Dyson index \cite{M-2004} accommodating system
symmetries ($\beta=1,\, 2,$ and $4$) whilst $\bar{\mathcal S}$ is the average
scattering matrix \cite{B-1997},
$
    \bar{\mathcal S}= V^\dagger \,{\rm diag}
    (\sqrt{1-\Gamma_j}
    )\, V
$, that characterises couplings between the cavity and the leads in
terms of tunnel probabilities \cite{R-06} $\Gamma_j$ of $j$-th mode
in the leads ($1\le j \le N$); the matrix $V$ is $V\in G(N)/G(N_{\rm
L}) \times G(N_{\rm R})$ where $G$ stands for orthogonal
($\beta=1$), unitary ($\beta=2$) or symplectic ($\beta=4$) group.

The above description becomes particularly simple for chaotic
cavities that coupled to the leads through ballistic point contacts
(``ideal'' leads, $\Gamma_j=1$). Indeed, uniformity of $P({\mathcal
S})$ over $G(N)$ implies that scattering matrices ${\mathcal S}$
belong \cite{BS-1990} to one of the three Dyson circular ensembles
\cite{M-2004} about which virtually everything is known.
Notwithstanding this remarkable simplicity, available analytic
results for statistics of electron transport are quite limited
\cite{B-1997,R-2008}. In particular, distribution functions of
conductance and shot noise power, as well as their higher order
cumulants, are largely unknown for an {\it arbitrary} number of
propagating modes, $N_{\rm L}$ and $N_{\rm R}$, and thus do not
catch up with existing experimental capabilities \cite{E-2001}.

In this Letter, we combine a stochastic version of the ${\mathcal
S}$-matrix approach with ideas of integrability
\cite{K-2002,OK-2007} to show that the problem of universal quantum
transport in chaotic cavities with broken time-reversal symmetry
($\beta=2$) is completely integrable. Although our theory applies
\cite{OK-2008} to a variety of transport observables, the further
discussion is purposely restricted to the statistics of Landauer
conductance. This will help us keep the presentation as transparent
as possible.

{\it Conductance distribution.}---In order to describe fluctuations
of the conductance $G = {\rm tr\,} ({\mathcal C}_1 {\mathcal S}
{\mathcal C}_2 {\mathcal S}^\dagger)$ in an adequate way, one needs
to know its entire distribution function. To determine the latter,
we define the moment generating function
\begin{eqnarray}
\label{c-iz}
    {\mathcal F}_n(z) = \left<
        \exp\left(
            - z G \right)
    \right>_{{\mathcal S}\in {\rm CUE}(2n+\nu)}
\end{eqnarray}
which, in accordance with the above discussion, involves averaging
over scattering matrices ${\mathcal S}\in {\rm CUE}(2n+\nu)$ drawn
from the Dyson circular unitary ensemble \cite{M-2004}. For the sake
of convenience, we have introduced the notation
$n=\min(N_{\rm L},N_{\rm R})$ and $\nu=|N_{\rm L} - N_{\rm R}|$ so
that the total number $N_{\rm L} + N_{\rm R}$ of propagating modes
in two leads equals $2n+\nu$.

While the averaging in Eq.~(\ref{c-iz}) can explicitly be performed
with the help of the Itzykson-Zuber formula \cite{IZ-1980}, a high
spectral degeneracy of the projection matrices ${\mathcal C}_1$ and
${\mathcal C}_2$ makes this calculation quite tedious. To avoid
unnecessary technical complications, it is beneficial to employ a
polar decomposition \cite{H-1963} of the scattering matrix. This
brings into play a set of $n$ transmission eigenvalues ${\bm T} =
(T_1,\cdots,T_n) \in (0,1)^n$ which characterise the conductance
\cite{LFLB-1957} in a particularly simple manner, $G({\bm T}) =
\sum_{j=1}^n T_j$.

The uniformity of the scattering ${\mathcal S}$-matrix distribution
gives rise to a nontrivial joint probability density function of
transmission eigenvalues in the form \cite{BM-1994,F-2006}
\begin{eqnarray}
\label{PnT}
    P_n({\bm T}) = c_n^{-1} \,\Delta_n^2({\bm T}) \prod_{j=1}^n
    T_j^\nu.
\end{eqnarray}
Here $\Delta_n({\bm T})=\prod_{j<k} (T_k-T_j)$ is the Vandermonde
determinant and $c_n$ is the normalisation constant \cite{M-2004}
\begin{eqnarray}
\label{nc}
    c_n = \prod_{j=0}^{n-1} \frac{\Gamma(j+2)\, \Gamma(j+\nu+1) \,\Gamma(j+1)}
    {\Gamma(j+\nu+n+1)}.
\end{eqnarray}
Let us stress that the description based on Eq. (\ref{PnT}) is
completely equivalent to the original, microscopically motivated
${\mathcal S}\in {\rm CUE}(2n+\nu)$ model.

Now the moment generating function can elegantly be calculated. A
close inspection of the integral
\begin{eqnarray}
\label{fnz}
    {\mathcal F}_n(z) =  c_n^{-1} \int_{(0,1)^n} \prod_{j=1}^n dT_j\, T_j^\nu \exp(-zT_j)\cdot \Delta_n^2({\bm T})
\end{eqnarray}
reveals that it admits the Hankel determinant representation
\cite{K-2002}
\begin{eqnarray}
\label{hd}
    {\mathcal F}_n(z) = \frac{n!}{c_n} \, {\rm det}\left[
        (-\partial_z)^{j+k}\, {\mathcal F}_1(z)
    \right]
\end{eqnarray}
with
\begin{equation}
\label{f1}
 {\mathcal F}_1(z) = \frac{(\nu+1)!}{z^{\nu+1}}
    \left(
            1 - e^{-z} \sum_{\ell=0}^\nu \frac{z^\ell}{\ell!}
    \right).
\end{equation}
In deriving Eqs.~(\ref{hd}) and (\ref{f1}) we have used the
Andr\'eief--de Bruijn integration formula \cite{A-1883}.

Equation (\ref{hd}), supplemented by the ``initial condition''
${\mathcal F}_0(z)=1$, has far-reaching consequences. Indeed, by
virtue of the Darboux theorem \cite{D-1972}, the infinite sequence
of the moment generating functions $({\mathcal F}_1,{\mathcal
F}_2,\cdots)$ obeys the Toda lattice equation $(n\ge 1)$
\begin{equation}
\label{TL}
    {\mathcal F}_n(z)\,{\mathcal F}_n^{\prime\prime}(z) - \left(
    {\mathcal F}_n^\prime(z)\right)^2 = {\rm var}_{n}(G)\,
    {\mathcal F}_{n-1}(z)\, {\mathcal F}_{n+1}(z),
\end{equation}
where ${\rm var}_{n}(G) = n(n+1)^{-1} (c_{n-1} c_{n+1}/ c_n^2)$ is
nothing but the conductance variance
\begin{equation}
\label{varG}
    {\rm var}_{n}(G) =  \frac{n^2 (n+\nu)^2}{(2n+\nu)^2
    [(2n+\nu)^2-1]}.
\end{equation}
Since ${\mathcal F}_n(z)$ is the Laplace transform of conductance
probability density $f_n(g)=\langle \delta (g - G) \rangle$, the
Toda lattice equation provides an exact solution \cite{R-09} to the
problem of conductance distribution in chaotic cavities with an
arbitrary number of channels in the leads. Equations (\ref{f1}) --
(\ref{varG}) represent the first main result of the Letter.

There exists yet another way to describe the conductance
distribution. Spotting that the moment generating function
${\mathcal F}_n(z)$ is essentially a Fredholm determinant
\cite{TW-1994} associated with a gap formation probability
\cite{M-2004} within the interval $(z,+\infty)$ in the spectrum of
an auxiliary $n \times n$ Laguerre unitary ensemble,
\begin{eqnarray}
    {\mathcal F}_n(z) \propto z^{-n(n+\nu)} \int_{(0,z)^n}
    \prod_{j=1}^n d\lambda_j \, \lambda_j^\nu \, e^{-\lambda_j}
    \cdot \Delta_n^2({\bm \lambda}),
\end{eqnarray}
one immediately derives \cite{TW-1994,FW-2002}:
\begin{eqnarray}
\label{FnP}
    {\mathcal F}_n(z) = \exp\left(
        \int_0^z dt \frac{\sigma_{\rm V}(t) - n(n+\nu)}{t}
    \right).
\end{eqnarray}
Here, $\sigma_{\rm V}(t)$ satisfies the Jimbo-Miwa-Okamoto form of
the Painlev\'e V equation \cite{JMMS-1980}
\begin{eqnarray}
\label{pv}
    (t \sigma_{\rm V}^{\prime\prime})^2 &+& [\sigma_{\rm V} - t \sigma_{\rm V}^\prime
    + 2 (\sigma_{\rm V}^\prime)^2 + (2n+\nu)\sigma_{\rm V}^\prime]^2 \nonumber\\
    &+& 4(\sigma_{\rm V}^\prime)^2 (\sigma_{\rm V}^\prime + n) (\sigma_{\rm V}^\prime+n+\nu)=0
\end{eqnarray}
subject to the boundary condition $\sigma_{\rm V}(t\rightarrow
0)\simeq n(n+\nu)$.

To the best of our knowledge, this is the first ever appearance of
Painlev\'e transcendents in problems of quantum transport. The
representation Eq.~(\ref{FnP}), being the second main result of the
Letter, opens a way for a nonperturbative calculation of conductance
cumulants.
\begin{widetext}
{\it Conductance cumulants.}---Our third main result is the bilinear
recurrence relation ($j \ge 2$)
\begin{equation}
\label{cumeq}
     [(2n+\nu)^2-j^2]\,(j+1) \kappa_{j+1} = 2 \sum_{\ell=0}^{j-1} (3\ell+1) (j-\ell)^2\left(\atop{j+1}{\ell+1}\right)
     \kappa_{\ell+1} \kappa_{j-\ell} - (2n+\nu) (2j-1)\,j \kappa_j - j(j-1)(j-2)\,
     \kappa_{j-1}
\end{equation}
\end{widetext}
for conductance cumulants $\{\kappa_j\}$. Taken together with the
initial conditions provided by the average conductance $\kappa_1 =
n(n+\nu)/(2n+\nu)$ and the conductance variance $\kappa_2 =
\kappa_1^2 /[(2n+\nu)^2-1]$, this recurrence efficiently generates
(previously unavailable) conductance cumulants of {\it any} given
order.

To prove Eq.~(\ref{cumeq}), we compare Eq.~(\ref{FnP}) with the
definition of the cumulant generating function
\begin{eqnarray}
    \log {\mathcal F}_n(z) = \sum_{j=1}^\infty \frac{(-1)^j}{j!} \,
    \kappa_j\, z^j
\end{eqnarray}
to deduce the remarkable identity
\begin{eqnarray}
\label{rel}
    \sigma_{\rm V}(z) = n(n+\nu) + \sum_{j=1}^\infty
    \frac{(-1)^j}{(j-1)!} \, \kappa_j\, z^j.
\end{eqnarray}
Substituting it back to Eq.~(\ref{pv}), we discover
Eq.~(\ref{cumeq}) as well as the above stated initial conditions.

{\it Large-$n$ limit of the theory.}---The nonpeturbative solution
Eq.~(\ref{cumeq}) has a drawback: it does not supply much desired
{\it explicit} dependence of conductance cumulants $\kappa_j$'s on
$j$. To probe the latter, we turn to the large-$n$ limit of the
recurrence Eq.~(\ref{cumeq}). For simplicity, the asymmetry
parameter $\nu$ will be set to zero.

Since, in the limit of a large number of propagating modes ($n\gg
1$), the conductance distribution is expected \cite{P-1989} to
follow the Gaussian law
\begin{eqnarray}
\label{Gd}
 f_n^{(0)}(g) = \frac{1}{\sqrt{2\pi \,{\rm var}_\infty(G)}}\,
 \exp\left(
    - \frac{(g-n/2)^2}{2\,{\rm var}_\infty(G)}
 \right)
\end{eqnarray}
with the average conductance ${\mathbb E}[G]=n/2$ and the
conductance variance ${\rm var}_\infty(G)=1/16$, it is natural to
seek a large-$n$ solution to Eq.~(\ref{cumeq}) in the form $\kappa_j
= (n/2)\delta_{j,1}+(1/16)\delta_{j,2} + \delta\kappa_j$, where
$\delta\kappa_{j}$ (with $j\ge 3$) account for deviations from the
Gaussian distribution. Next, we put forward the large-$n$ ansatz
\begin{eqnarray}
\label{1-over-n} \delta\kappa_j =
\frac{1}{n^j}\sum_{m=0}^\infty\frac{a_m(j)}{n^{m}}
\end{eqnarray}
which, after its substitution into the recurrence, yields the
explicit formula
\begin{equation}
\label{a0}
\delta\kappa_{2j} = \frac{1}{4} \frac{(2j-1)!}{(4n)^{2j}}
\left[
    1+ \frac{j(3j^2-1)}{8 n^2} +{\cal O}\left(\frac{1}{n^4}\right)
\right].
\end{equation}
All odd order cumulants vanish identically.

Interestingly, Eq.~(\ref{a0}) makes it possible to analytically
study a deviation of conductance distribution $f_n(g)$ from the
Gaussian law $f_n^{(0)}(g)$. The Gram-Charlier expansion
\begin{eqnarray}
\label{GCh}
     f_n(g) = \exp\left(
        \sum_{j=1}^\infty \frac{\delta\kappa_j}{j!} (-\partial_g)^j
     \right) \, f_n^{(0)}(g)
\end{eqnarray}
is the key. As soon as $|\partial_g \log f_n^{(0)}(g)| \sim n$, the
operator in the exponent is dominated by the $m=0$ term in
Eq.~(\ref{1-over-n}). This observation reduces Eq.~(\ref{GCh}) to
\begin{equation}
\label{int-f}
    f_n(g) = \frac{2 n^{1/4}}{\Gamma(1/8)}\sqrt{\frac{2}{\pi}}
     \int_0^\infty \frac{d\lambda\, e^{-n^2\lambda}}{\lambda^{7/8}\sqrt{1+2\lambda}}
      \exp\left(
    -\frac{2 n^2\,\eta^2}{1+2\lambda}
    \right).
\end{equation}
Here, $\eta$ is the rescaled conductance $\eta = 2(g/n)-1$.

Equation (\ref{int-f}) is particularly suitable for the asymptotic
analysis. Performed with a logarithmic accuracy, it brings:
\begin{equation}
\label{f-26}
    \log \, f_n(g) \sim \left\{
                          \begin{array}{ll}
                            -2 n^2\,\eta^2, &\;\;\; |\eta| < \displaystyle \frac{1}{ 2} \\
                            -2 n^2 \left(|\eta| -\displaystyle \frac{1}{4}\right)
                            -\displaystyle \frac{3}{4} \log n, & \;\;\;\displaystyle \frac{1}{2} < |\eta| < 1
                          \end{array}
                        \right.
\end{equation}
This result shows that the Gaussian approximation for the
conductance distribution is only valid for \linebreak $|g-n/2|<n/4$.
Away from this region, the conductance distribution exhibits long
tails described by the exponential rather than the Gaussian law.
Finally, it is straightforward to derive from the Toda lattice
Eq.~(\ref{TL}) that, in the vicinity $|g-g_*| \le 1$ of the edges
\cite{R-09} $g_*=0$ and $g_*=n$, the conductance distribution
exhibits even slower, power-law decay \cite{MB-1999,R-2008}
\begin{equation}
\label{p-law}
    \log f_n(g) \sim (n^2-1) \log \left(2 \,|\eta-\eta_*|\right) - \frac{n^2}{2} + \frac{1}{12}\log
    n
\end{equation}
with $\eta_*=\pm 1$.

{\it Conclusions.}---We have shown that a marriage between the
scattering ${\mathcal S}$-matrix approach and the theory of
integrable systems brings out an efficient formalism tailor-made to
analysis of the universal aspects of quantum transport in chaotic
systems with broken time-reversal symmetry. Having chosen the
paradigmatic problem of conductance fluctuations in chaotic cavities
with ideal leads as an illustrative example, we determined the
cumulants of conductance as well as its distribution exactly for any
given number of propagating modes in the leads. It should be
stressed that the ideas presented in the Letter can equally be
utilised \cite{OK-2008} to describe statistical properties of the
shot noise power and the dynamics of charge transfer.

Certainly, more effort is needed to accomplish integrable theory of
the universal quantum transport. Extension of the formalism
presented to the $\beta=1$ and $4$ symmetry classes and waiving the
uniformity of the ${\mathcal S}$-matrix distribution are the two
most challenging problems whose solution is very much called for.

This work was supported by the Israel Science Foundation through the
grant No 286/04.

{\it Note added.}---Recently, we learnt about the paper by M.~Novaes
\cite{N-2008} who noticed that the $n$--th {\it moment} of
conductance can nonperturbatively be calculated by using the
machinery of hypergeometric functions of matrix argument. Neither
Toda lattice nor Painlev\'e V representations for the conductance
distribution surfaced there.

%\vspace{-0.5cm}


\begin{references}
\bibitem{R-02} Energy averaging is performed over such a small energy window near the Fermi
    energy that keeps the classical dynamics essentially unchanged.
\bibitem{R-00} The Ehrenfest time $\tau_{\rm E}\simeq \lambda^{-1} \log (W/\lambda_F)$
    is determined by the Lyapunov exponent $\lambda$ of
    chaotic classical dynamics, the Fermi wavelength $\lambda_F$, and the lead widths $W$.
    The mean dwell time $\tau_{\rm D}\simeq A/(W v_F)$, where $A$ is
    the area of the cavity.
\bibitem{R-03} The notion of universality should be taken with some
    care since transport observables will generically depend on
    non-universal couplings between the cavity and the leads, see Eqs.~(\ref{ham}) -- (\ref{pk}).
\bibitem{R-06} The tunnel probabilities $\Gamma_j = 4{\tilde w}_j/(1+{\tilde w}_j)^2$ are determined
    by the eigenvalues ${\tilde w}_j$ of the $M\rightarrow \infty$ matrix
    $(\pi^2/M\Delta)\,{\mathcal W}^\dagger {\mathcal W}$, where $\Delta$ is the mean level spacing.
    The average electron dwell time $\tau_{\rm D} = (2\pi \hbar/\Delta)(\sum_{j=1}^N \Gamma_j)^{-1}$.
\bibitem{R-09} Due to a specific form of ${\mathcal F}_1(z)$, a calculation of the inverse Laplace transform of
    ${\mathcal F}_n(z)$ is straightforward. The resulting
    probability density function $f_n(g)$ is seen to be
    a nonanalytic function with a finite support $g\in(0,n)$.
\end{references}
\end{document}